\documentstyle[12pt,fullpage,psfig]{article}
\def\beq{\begin{equation}}
\def\eeq{\end{equation}}

\begin{document}
\begin{center}
{\Large{\bf Some Results And A Conjecture For Manna's Stochastic
Sandpile Model}}
\\[2cm]

{\large{\bf Deepak Dhar}}\\
Department of Theoretical Physics, \\ 
Tata Institute of Fundamental Research, \\ 
Homi Bhabha Road, Mumbai 400~005, INDIA\\ [1cm]
\underbar{Abstract}
\end{center}

We present some analytical results for the stochastic sandpile model
studied earlier by Manna. In this model, the operators corresponding to
particle addition at different sites commute. The eigenvalues of operators
satisfy a system of coupled polynomial equations. For an $L \times L$
square, we construct a nontrivial toppling invariant, and hence a ladder
operator which acting on eigenvectors of the evolution operator gives new
eigenvectors with different eigenvalues. For periodic boundary conditions
in one direction, one more toppling invariant can be constructed.  We show
that there are many forbidden subconfigurations, and only an exponentially
small fraction of all stable configurations are recurrent. We obtain
rigorous lower and upper bounds for the minimum number of particles in a
recurrent configuration, and conjecture a formula for its exact value for
finite-size rectangles.

PACS nos: 05.20.Ln, 02.50.Ga, 02.10.Pk, 81.05.Rm

\section{ Introduction} 

  Many cellular automaton `sandpile' models have been studied in recent
years as simple theoretical models of self-organized criticality. The
prototype of these models is the Abelian sandpile model (ASM), proposed by
Bak et al in 1987 \cite{BTW1,BTW2}. Many properties of this model can be
determined analytically using the fact that the operators corresponding
particle addition in this model generate an abelian group, and the model
is equivalent to the $ q \rightarrow 0$ limit of the Potts model
\cite{dd99}. Several variations of the ASM have been studied in the past
with a view to understand the parameters that determine the different
universality classes of self-organized critical behavior
\cite{knwz,manna1,zhang}. These include models in which particle transfer
is directed \cite{ddrr}, or models in which the toppling condition or the
number of sandgrains transferred depends on the local slope rather than
local height. In this respect, it has been realized that stochasticity in
toppling rules can lead to different critical behavior than models with
deterministic toppling rules \cite{manna2,bosa, benhur}.
 
 As a physical motivation for the study of such models, we
note that stochasticity in the toppling rules may be taken as a simple
phenomenological attempt to take into account in a theoretical description
the variation in shape and smoothness of different grains in real granular
media. It has been argued that a stochastic one-dimensional sandpile model
seems to describe well the behavior of experimental results on avalanches
in ricepiles \cite{ricepile}.

In this paper, we study a class of stochastic sandpile models, the
prototype of which is the so-called Manna model \cite{manna2}. It was
recently proved that this model also shows the abelian property, and is a
special case of the more general Abelian Distibuted Processors (ADP) model
\cite{dd99}. In the following, we shall use the terms deterministic ASM
(DASM) and stochastic ASM (SASM) if we need to distinguish between
these classes of models.  We will show that some of the
analytical techniques of DASM are also useful in studying the SASM.

The plan of this paper is as follows: In section 2, we recapitulate the
definition of the Manna model. We study this model in two dimensions on a
square lattice, but most of the treatment can be easily generalized to
higher dimensions.  In section 3, we describe the abelian ring structure
of the algebra generated by the particle addition operators. The
simultaneous eigenvalues of these operators satisfy a system of coupled
quadratic equations. In section 4, we show that for the case of an $L
\times M$ rectangle, if $L = M$, there is a nontrivial toppling invariant.
We use this to construct a ladder operator that gives $(L+1)$ simultaneous
eigenvectors of the particle addition operators from one such eigenvector.
For periodic boundary conditions in the y-direction, and $M=L+1$, we can
construct one more toppling invariant, and this raises the number of
eigenvectors generated from one vector to $ M^2$. In section 5, we prove
the existence of forbidden sub-configurations (FSC's) in the model. We
describe an algorithm to construct the set of minimal FSC's. In section 6,
we study the recurrent configurations of the model which have minimum
number of sandgrains. We prove upper and lower bounds on this number for a
finite rectangle, and conjecture that the upper bound always coincides
with the exact value.

\section{Definition of the Model}

 We consider a two-dimensional square lattice. The sites of the lattice
will be labelled by their integer Euclidean coordinates $(x,y), ~~1 \leq x
\leq L, ~~1 \leq y \leq M$. At each site $(x,y)$, there is a non-negative
integer $z_{x,y}$ called the height of the pile. In addition, at each
site,we assume that we have a pseudo-random-number generator (PRNG), which
gives a number lying between $0$ and $1$ on each request using a
deterministic algorithm. The $n$-th output from a PRNG depends only on its
initial setting (seed), and $n$, and not in any way on the evolution of
the system, or status of other PRNG's. Equivalently, we can think of the
PRNG's as an infinite stack of independent random numbers at each site.

If the height at any site does not exceed $1$, the configuration is said
to be stable.  If the system is in a stable configuration, we choose a
site at random and add a grain of sand there. as a result of the addition,
height at that site increases by 1.  If this makes the site unstable, we
relax the system using the relaxation rule given below. Once all unstable
sites are relaxed, we add another particle. We shall assume that the
probability of addition is nonzero for all sites. It follows from the
general theory of ASM's that the steady state of the system is then
independent of the precise choice of these probabilities.

The relaxation rule is as follows:  Any site whose height exceeds $1$,
relaxes by toppling, and throwing two grains out. Where these grains go
is decided by drawing a new random number from the local PRNG. In terms of
stacks, we pop up a number from the local stack. If the number is
less that $1/2$, one particle is transferred to each of the two horizontal
neighbors of the site. Else, one particle is transferred to each of the
two vertical neighbors.  If the transfer renders some other sites
unstable, these are toppled in turn, until a stable configuration is
reached. Sandgrains can leave the system if a toppling occurs at a
boundary site.

\section{The Abelian Ring of Operators}

This model as a special case of the more general Abelian Distributed
Processors Model (ADP) \cite{dd99}, and the abelian character of the model
becomes obvious.  We define particle addition operators ${\bf A}_{x,y}$
which acting on a stable configurations ${\cal C}$, gives the stable
configuration obtained after adding a particle at $(x,y)$, and relaxing
the system. Here the stable configurations are specified by the heights
$\{z_{x,y}\}$, and the state of the PRNG at each site (equivalently, a
stack of random numbers at each site). It is easy to see that if we start
with a configuration with two or more  unstable sites, we get the same
configuration, independent of the order in which
the unstable sites are relaxed. Using this property repeatedly, it follows
that

\beq 
[{\bf A}_{x,y},{\bf A}_{x',y'}]=0,~~{\rm for~ all~ sites}~~
(x,y)~{\rm and~} (x',y'). 
\eeq
 
We may now assume that the internal state of these PRNG's as inaccessible
to the outside observer, and that there are no observable correlations
between different outputs of the PRNG's ( i.e. they are `good' PRNG's).
Then the deterministic evolution of avalanches in the (sandpile heights+
PRNGs) system may be described equally well as a {\it Markovian
relaxation} of the subsystem described in terms of sandpile heights alone.

Given a particular state of the PRNG's at each site, addition of two
particles at different sites gives same final stable configuration,
independent of the order of topplings. A different state of PRNG's can
lead a different final state, but still independent of order of topplings.
If we do not know the state of the PRNG's, we have to average over their
different possible states. Then, the final state is not fixed, only the
probabilities of different outcomes can be given.

 In Manna's original formulation of the model, the transfer direction was
decided randomly at each toppling. The advantage of the specific
implementation used here with a different PRNG at each site makes the fact
that abelian character of the model obvious. The abelian property, once
established, is clearly independent of the details of implementation of
the algorithm. So, one could as well use the same PRNG to decide the
direction of particle transfer in topplings at different sites.

We consider the $2^{LM}$ -dimensional vector space $\cal V$, whose basis
vectors $|{\cal Z}>$ are labelled by the stable height configurations
${\cal Z} \equiv \{z_{x,y}\}$.  If the probability that at time $t$ the
configuration of the sandpile is ${\cal Z}$ is denoted by $Prob({\cal
Z},t)$, then the state of the system can be represented by a vector
$|P(t)> \in \cal V$

\beq
|P(t)> = \sum_{{\cal Z}} Prob({\cal Z},t)|{\cal Z}>
\eeq

We define particle addition operators ${\bf a}_{x,y}$ as linear operators
acting on $\cal V$ as follows: for all configurations ${\cal Z}$,
${\bf a}_{x,y}|{\cal Z}>$ gives the probability vector of the system
obtained by taking the system in the stable configuration ${\cal Z}$,
adding a particle at site $(x,y)$, and relaxing the system until a stable
configuration is reached.  For stochastic toppling rules, the resulting
state is not necessarily a basis vector corresponding to a unique
configuration, but a linear combination of different basis vectors. Note
that action of any of these operators on a given configuration gives a
unique probability state vector, and action of these $\{{\bf a}_{x,y}\}$'s
is deterministic, in the same way as evolution of the wave-function in
quantum mechanics is deterministic.  The Abelian property of
the Manna model means that {\it the probabilities of different final
stable configurations are independent of the order of topplings and
particle additions}.

The commutativity of the operators $\{{\bf A(x,y)}\}$ implies that
the operators $\{{\bf a}_{x,y}\}$ also commute with each other:

\beq
[{\bf a}_{x,y},{\bf a}_{x',y'}]=0,{\rm ~~for~ all~ sites~} (x,y)~ {\rm 
and}~ (x',y').
\eeq

This abelian property simplifies the analysis of the model
considerably. There are additional relations satisfied by the
operators $\{{\bf a}_{x,y}\}$. Adding two particles at a site will
cause it to topple, whatever the initial configuration, and then
with equal probability two particles are transferred either horizontally,
or vertically.
This implies that the operators $\{{\bf a}_{x,y}\}$  satisfy the
equations

\beq
{\bf a}_{x,y}^2 = \frac{1}{2}({\bf a}_{x-1,y}{\bf a}_{x+1,y} + {\bf
a}_{x,y-1}{\bf a}_{x,y+1}),{\rm ~~ for~~ all ~~} x,~y,~ ( 1\leq x \leq L,
1 \leq
y \leq M).
\eeq

In writing these equations, we have assumed the boundary conditions

\beq
{\bf a}_{0,y} = {\bf a}_{L+1 ,y} =1,  {\rm ~~for }~~ 1 \leq y \leq M
\eeq
\beq
 {\bf a}_{x,0} = {\bf a}_{x,M+1} =1,  {\rm ~~for} ~~ 1 \leq x \leq L
\eeq

If we assume periodic boundary conditions in the y-direction, the
Eqs.(6) are replaced by the equations

\beq
{\bf a}_{x,y}={\bf a}_{x, y+M},{\rm ~~for ~all}~x, ~1 \leq x \leq L,{\rm
~all} ~~y.
\eeq

Any vector ${\bf v} \in {\cal V}$ can be expressed as $P({\bf
v})|\Phi>$, where $P({\bf v})$ is a polynomial in $\{{\bf
a}_{x,y}\}$, with the highest power of any ${\bf a}_{x,y}$ being
$1$, and $|\Phi>$ is the vector corresponding to configuration
with all sites empty. We can multiply two such polynomials $P_1$
and $P_2$, and reduce the highest power of operators ${\bf
a}_{x,y}$ using the reduction rules Eq.(4). 

We may think of the operators $\{{\bf a}_{x,y}\}$ as abstract symbols, and
study the algebra of polynomials in these variables defined by the usual
addition, and multiplication. Using the reduction rules (4), the maximum
degree of any symbol in the polynomials can be reduced to $1$. Clearly,
the set of all such polynomials forms a commutative ring.

It is instructive to compare this situation with that of the DASM. In the
latter case also, there are reduction rules which reduce large powers of
abelian operators $\{{\bf a}_{x,y}\}$ to lower ones, but the expression on
the right-hand side of Eq. (4) consists of a single term. Then the
operators define a finite-dimensional abelian semi-group, and if we
restrict to the set of recurrent states, we get an abelian group with an
well-defined multiplicative inverse \cite{dd99}. In
the present case, the reduction rules give us not a single term, but a
linear combination of terms. To form an algebra closed under addition and
multiplication, we have to include all linear combinations of terms. Thus,
the number of elements of the algebra is infinite.

 The operators $\{{\bf a}_{x,y}\}$ have an obvious representation as $
2^{LM} \times 2^{LM}$ matrices ( defined by their action on basis vectors
of $\cal V$). These matrices are in general not diagonalizable, only
reducible to the Jordan canonical form, and there may not be eigenvectors
corresponding to each repeated eigenvalue.  For the DASM, all transient
states correspond to the zero eigenvalue of
the operators $\{{\bf a}_{x,y}\}$, and all other eigenstates are of
modulus $1$. For the SASM, there may be a set of
configurations such that if the system is any one of them, with a finite
probability, it makes a transition within the same set, and with a finite
probability it gets out. If the system can never get back to these states,
these states are transient, but there need not be any zero eigenvalue of
the operators $\{{\bf a}_{x,y}\}$ corresponding to such transient states.  
Also, amongst the recurrent states, one can have two different initial
probability vectors that yield the same resultant vector. Thus, there can
be eigenvectors of zero eigenvalue in the recurrent sector, and in
general, the inverse operator
${\bf a}_{x,y}^{-1}$ need not exist, even if we restrict ourselves to the
set of recurrent configurations of the model.

Consider one of the operators, say ${\bf a}_{1,1}$. Let $a_{1,1}$ be one
of the eigenvalues of ${\bf a}_{1,1}$. Let ${\cal V}_1$ be the subspace of
${\cal V}$ spanned by the (right) eigenvectors of ${\bf a}_{1,1}$
corresponding to the eigenvalue $a_{1,1}$. There is at least one such
eigenvector, so ${\cal V}_1$ is non-null. We pick one of the other
addition operators, say ${\bf a}_{1,2}$. From the fact that ${\bf
a}_{1,2}$ commutes with ${\bf a}_{1,1}$, it immediately follows that ${\bf
a}_{1,2}$ acting on any vector in the subspace ${\cal V}_1$ leaves it
within the same subspace. Diagonalizing ${\bf a}_{1,2}$ within this
subspace, we construct a possibly smaller, but still non-null, subspace
${\cal V}_2$ which is spanned by simultaneous eigenvectors of ${\bf
a}_{1,1}$ and ${\bf a}_{1,2}$ with eigenvalues $a_{1,1}$ and $a_{1,2}$.
Repeating this argument with other operators, we can construct vectors
which are simultaneous eigenvectors of all the $\{{\bf a}_{x,y}\}$. Let
$|\Psi>$ be such a vector, with

\beq
{\bf a}_{x,y}|\Psi> =  a_{x,y} |\Psi>, {\rm ~for ~all} ~(x,y).
\eeq

Then the eigenvalues $\{a_{x,y}\}$ satisfy the coupled set of
quadratic equations

\beq
a_{x,y}^2 = \frac{1}{2}( a_{x-1,y} a_{x+1,y} + a_{x,y-1}  a_{x,y+1}),~~
{\rm for~~ all ~~} x,~y;~~  1\leq x \leq L, 1 \leq
y \leq M.
\eeq

These are $LM$ quadratic equations in the $LM$ complex variables
$\{a_{x,y}\}$. In general, they have $2^{LM}$ solutions. It seems
difficult to write down the general solution of these equations
explicitly. However, it is easy to see that all $a_{x,y}=1$ is certainly a
solution of these equations. The corresponding eigenvector $|\Psi>$ gives
the steady state. Unlike the DASM case, here we are not able to write down
explicitly the probabilities of different recurrent configurations in the
steady state.

\section{Toppling Invariants}

Toppling invariants have been found to be very useful in characterizing
the structure of recurrent states in the DASM case \cite{drsv}. Consider a
DASM on N sites, whose configuration is specified by the heights $\{z_i\},
i=1,2 \ldots N$, and the toppling matrix is $\Delta$. For any arbitrary
configuration ${\cal Z} \equiv \{ z_i\}$, a toppling invariant $I$ is
defined as a integer function linear in the height variables by the
equation

\beq
I({\cal Z}) = \sum_{i} g_i z_i, ~~~~( mod ~~d).
\eeq
where $d$ is some integer, and $g_i$ are integers chosen so that $I$ does
not change under toppling at any of the sites $j$. The $N$ different
choices of $j$ give $N$ conditions to be satisfied by the as yet
undetermined coefficients $\{g_i\}$. For suitable choice of $d$, these
equations have a non-trivial solution, and in fact one can find a minimal
set of toppling invariants in terms of the Smith normal form of the matrix
$\Delta$ \cite{drsv}.

For the present model, we define a toppling invariant $I_1$ by the
equation 

\beq
I_1({\cal Z}) = \sum_{x,y} g(x,y) ~z_{x,y}, ~~~~( mod~~ d).
\eeq
For this to be  unchanged if there is a toppling at $(x,y)$, we must have

\beq
2 g(x,y) = g(x-1,y) + g(x+1,y), ~~~~~ (mod ~~d).
\eeq
and

\beq
2 g(x,y) = g(x,y-1) + g(x,y+1), ~~~~~ (mod ~~d).
\eeq
These are $2LM$ equations for the $LM+1$ unknown coefficients $g(x,y)$ and
$d$. In general, we would not expect any nontrivial solution.
Interestingly, for  the special case $L = M$, these equations have a
simple nontrivial solution for $d=L+1$:

\beq
g(x,y) = xy
\eeq
 
We show below that toppling invariants can
also be used as ladder operators, which acting on an eigenvector of the
evolution operator, gives rise to another eigenvector of the same with a
different eigenvalue. For the DASM, this is not very important, as the
evolution operator can be diagonalized directly, and all the eigenvectors
are easily determined. For the SASM, the eigenvectors are not easily
determined, and the fact that toppling invariants can be used to construct
ladder operators provides useful information. 
    
We define an operator ${\bf S}$, diagonal in the configuration basis by
the equation

\beq
{\bf S} = exp ( \frac {-2 \pi i I_1}{L+1}) 
\eeq
Clearly, ${\bf S}$ is also invariant under toppling. Simple algebra then
gives

\beq
{\bf S}^{-1}~ {\bf a}_{x,y}~ {\bf S} =  exp ( \frac {2 \pi i x
y}{L+1})~{\bf a}_{x,y}
\eeq
Let $|\Psi>$ is a simultaneous eigenvector of the operators $\{{\bf
a}_{x,y}\}$, with eigenvalues $\{a_{x,y}\}$, $1 \leq x \leq L, 1 \leq y
\leq M$. Then, it follows that ${\bf S}|\Psi>$ is also a simultaneous
eigenvector of all these operators with eigenvalues $\{\tilde{a}_{x,y}\}$,
where

\beq
\tilde{a}_{x,y} = a_{x,y}~ exp[ \frac {2 \pi i x y}{L+1}].
\eeq
If we use periodic boundary conditions in the y-direction, and $L= M-1$, 
we can define another toppling invariant $I_2$ 

\beq
I_2({\cal Z}) = \sum_{x,y} x z_{x,y}, ~~~~ mod M.
\eeq

Thus, for a lattice of size $L \times (L+1)$, we have two independent
toppling invariants $I_1$ and $I_2$. More generally, if $M \neq L+1$, and
the greatest common divisor of $M$ and $L+1$ is $R$, then both $I_1$ and
$I_2$ are invariant modulo $R$.

 Using these symmtries together, we can construct a set of $M^2$ solutions of
the coupled eigenvalue equations (9) when $ M=L+1$, for any solution
$\{a_{x,y}\}$. These solutions are parameterized by two integers $k$ and
$r$. It is easily checked that if $\{a_{x,y}\}$ is a solution of Eqs.(9),
then so is $\{{\tilde a}_{x,y}\}$, where

\beq
\tilde{a}_{x,y} = a_{x,y} exp[ 2 \pi i (k+ r y) x/M ],~~0 \leq k < M, ~0
\leq r < M
\eeq

We can group configuration having the same values of $I_1, I_2$ into same
equivalence class. There are $M^2$ such equivalence classes. Clearly, no
value of $I_1$ or $I_2$ has any preferred status, and so, in the steady
state, each of these equivalence classes must have equal weight.  Thus,
the toppling invariants provide a partial characterization of the
probabilities of different configuration in the steady state. Also, it is
easy to see that  all configuration obtained by adding a particle at
$(x,y)$ to  a configuration belongs to the equivalence class of
$I_1= i, I_2 =j$, would belong to the equivalence class with $I_1= i+ x y,
I_2 = j + x$.

\section{ Minimal Forbidden Subconfigurations}

Determining the probabilities of different stable configurations in the
steady state is not easy. However, as a first step towards such a
characterization, we show below that most of the $2^{LM}$ stable
configurations do not occur at all in the steady state, and the number of
recurrent configurations of the sandpile is only an exponentially small
fraction of all stable configurations.  We do this by showing that there
are many local constraints on the allowed heights $\{z_{x,y}\}$ in
recurrent configurations.

\begin{figure}
\begin{center}
\leavevmode
\psfig{figure=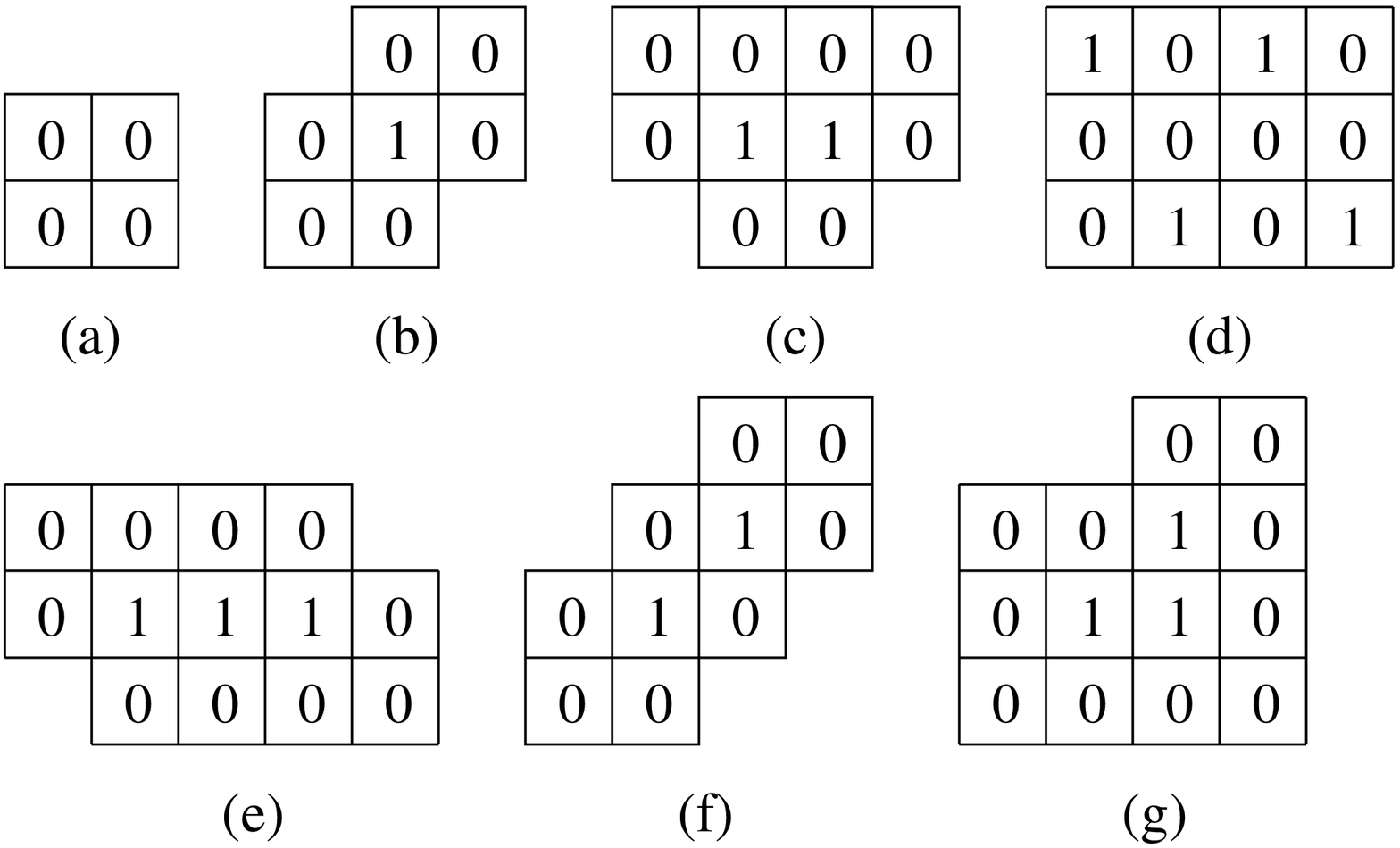,width=12cm,angle=0}
\caption{ Some minimal forbidden subconfigurations of the Manna
Model}
\end{center}
\end{figure}

The simplest of such constraints is that in any recurrent configuration,
any four neighboring sites forming a $2 \times 2$ square cannot be all
unoccupied. Thus the local structure shown in Fig. 1a cannot be found
anywhere in a recurrent configuration. We shall call such structures
forbidden subconfigurations (FSC's). In fact, we shall show that an FSC
cannot even  in an intermediate unstable configuration, if we start with a
recurrent configuration.

The proof  that Fig. 1a is an FSC is similar to the proof used in the
DASM case \cite{dd90}: We start with an initial configuration of the
sandpile in which there is no such subconfiguration. Now we evolve the
sandpile by adding particles at sites, and toppling any sites which become
unstable. We would like to prove that at no stage, in the intervening
stable or unstable configurations, would we be able to generate the
subconfiguration shown in Fig. 1a, whatever the sequence of particle
additions, and whatever choice we make for each toppling (horizontal or
vertical transfer).

To prove, we assume the contrary. Let us treat each particle addition or
toppling as a seperate relaxation step. Let the FSC appears for the first
time after the $T$-th step. Then, after step $T-1$, there was no FSC at
these sites, and the height at one of these sites must have decreased in
the $T$-th step. This implies that the $T$-th step was a toppling at one
of the four sites in the $2 \times 2$ square. But this toppling, whether
transfer was horizontal or vertical, would send one particle to one of the
neighbors in the square. Since the minimum height at the neighbor is zero
at time $T-1$, its height at the $T$-th step must be at least $1$, which
contradicts the assumption that the height is $0$ at all the four sites at
the $T$-th step. Hence, no such $T$ can be found, and the assumption is
not possible.

We can prove that other configurations shown in Fig. 1 are also FSC's
similarly. For example, to show that subconfiguration shown in Fig. 1b is
an FSC, we again note that if this subconfiguration is generated for the
first time at the $T$-th step, then the previous step must be a toppling
at one
of the sites in the FSC. That implies that at the $(T-1)$-th step, the
height at
one of the neighboring sites was negative, or that the central site was of
height zero . But the latter case would lead to an configuration shown
Fig. 1a, already proved to be a FSC.

If a subconfiguration cannot be generated on a small lattice, it cannot be
generated on a bigger lattice either, as the effect of the rest of the
lattice on a small part can always be reproduced by particles added at the
boundaries of the of the part at appropriate points in space and time.
Thus, every transient configuration of a small lattice is an FSC on a
bigger lattice.

Clearly, if ${\bf F}$ is an FSC, so is any bigger set containing ${\bf F}$
as a subset. We may therefore, without any loss, restrict our discussion
to FSC's such that any proper subset of the sites does not form an FSC.
Such FSC's will be called minimal FSC's. 

To form a list of all minimal FSC's, one can start by examining the
transient configurations on small lattices. It is easily seen that all
configurations are recurrent if $M$ or $L = 1$, and that Fig. 1a gives the
only transient configuration on a $2 \times 2$ lattice. One can check that
going to lattices $2 \times M$ with $M > 2$ does not generate any new
minimal FSC. We then enumerate all the transient configurations of the $3
\times 3$ lattice. Some of these are disallowed by the existence of FSC's
of smaller sizes. From the remaining transient configurations, we can get
a reduced set of minimal FSC's by replacing the FSC by a smaller set with
fewer sites wherever possible. This gives us a new minimal FSC shown in
Fig. 1b ( and another FSC related to this by reflection symmetry). Going
to bigger lattices, we can continue to augment our list of FSC's, with
larger and larger FSC's. At each stage, we can be sure that all minimal
FSC's having fewer than a specific number of sites have been enumerated.
Some of the minimal FSC's constructed in this way are shown in Fig. 1 b-g.  
If we confined our attention to FSC's on stable configurations only, and
not insist that they be forbidden in unstable states also, these
configurations could have been made smaller by omitting all sites of
height $1$ in the figure.

The existence of FSC's implies local inequality contraints in the height
variables that must be satisfied at all sites in recurrent states. This
implies that the fraction of the $2^{LM}$ stable configurations that are
recurrent must decrease to zero for large lattices ( exponentially fast in
the number of sites in the lattice).

The procedure to construct FSC's is rather tedious, and not very
illuminating as the number of minimal FSC's is infinite. It seems
desirable to have a simple algorithm which can identify FSC's. For the
DASM, such an algorithm is rather simple: it is known as the burning
test \cite{burning},
or the script test for some asymmetric graphs \cite{speer}.

 A simple variation of the burning test seems to work well for a large
majority of configurations: starting with the given configuration, one
recursively burns any site $i$ for which the height $z_i \geq
Min(v_i,h_i)$, where $v_i$ and $h_i$  are the number of vertical and
horizontal {\it unburnt} neighbors of $i$ respectively. If eventually all
sites are
burnt, the configuration passes this burning test, else not.

Some not-very-systematic checking on small lattices showed that if a
configuration passes this modified burning test, it is usually recurrent,
and if it fails, it is transient. However, there are a few instances where
this conclusion is incorrect. For example, the FSC shown in Fig. 1d burns
completely, and hence should have been recurrent according to this test.
One can also find configurations that do not burn according to this test,
but are recurrent. An example is given in Fig. 2.  It is possible that a
small change in the burning condition in this test would be make it work
without any error. Finding such a refinement remains an open problem at
present.

\begin{figure}
\begin{center}
\leavevmode
\psfig{figure=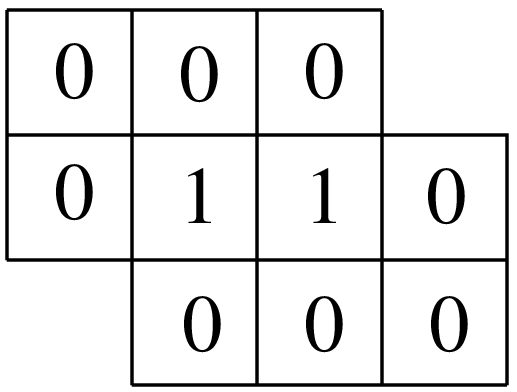,width=8cm,angle=0}
\caption{ A subconfiguration that does not burn in the modied burning
test, but is not an FSC.} 
\end{center}
\end{figure}

We note that the existence of FSC's is a special feature of these  
toppling rules. If we had chosen the particle transfer to be in the 
north-and-east or south-and-west directions with equal probability, then
it is easy to see that starting from any configuration, we can get to the
configuration with all sites empty, and hence all stable configurations
are recurrent.

\section{Minimal Sand Configurations}

Let ${\cal R}$ be a recurrent configuration of sandpile on a
rectangle of size $\ell \times m$. We denote
the number of sandgrains in ${\cal R}$ by $N({\cal R})$. The minimum
value of $N({\cal R})$ over all recurrent configurations on the lattice
will
be denoted by $N_{min}(\ell,m)$. In this section we study how
$N_{min}(\ell,m)$ depends on $\ell$ and $m$. For large $\ell, m$, this
should vary linearly with the number of sites in the lattice, and we
define
the minimum possible density of occupied sites in a recurrent configuration as 
$\rho_{min}$. Clearly,

\beq
\rho_{min} = Lim_{\ell, m \rightarrow \infty} \frac
{N_{min}(\ell,m)}{\ell m}.
\eeq

We have already seen that four adjacent sites forming a $2 \times 2$
square, if all unoccupied, form an FSC, and hence each such group of
four sites must have at least one occupied site. For large $\ell,m$,
we can divide the space into non-overlapping $2 \times 2$ squares,
and so we see that
\beq
\rho_{min} \geq \frac{1}{4}
\eeq

\begin{figure}
\begin{center}
\leavevmode
\psfig{figure=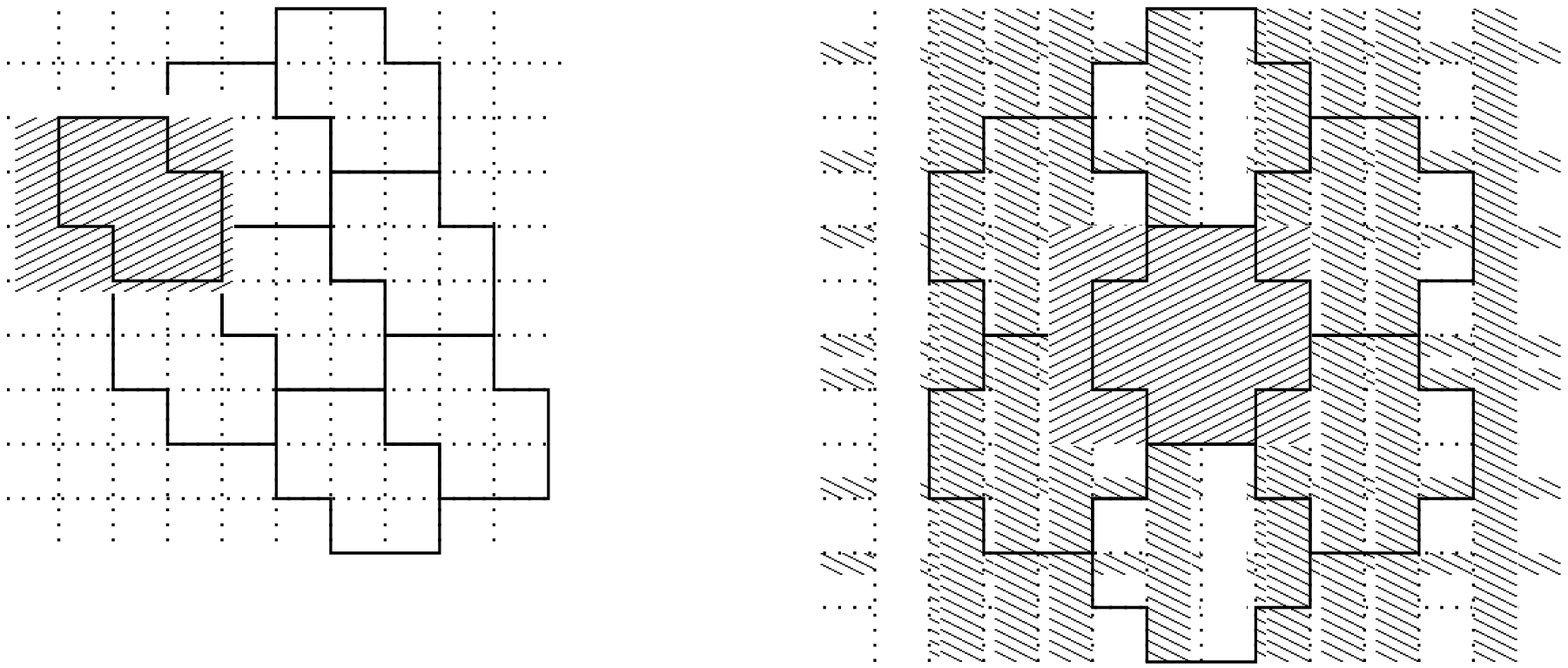,width=14cm,angle=0}
\caption{ Tiling of the plane using tiles of 7 and 12 sites.} 
\end{center}
\end{figure}

One can get better bounds on $\rho_{min}$ using bigger tiling units. In 
Fig. 3, we have shown the tilings using tiles using $7$ and $12$ sites
each. For the 7-site tile, it is easy to see that we get an FSC within
the tile, unless the number of occupied sites within the tile is at
least $2$. This implies that $\rho_{min} \geq \frac{2}{7}$. For the
$12-$ sited tiling, the minimum number of occupied site per tile can
easily be checked to be $4$. This gives us an improved bound 
$\rho_{min} \geq \frac{1}{3}$.

It is straight-forward to extend this procedure to larger tiles. For
any finite tile, there are only a finite number of configurations, and
by exhaustive enumeration, one can find  the largest
number $r$, such that
all configurations with less than $r$ occupied sites necessarily contain
an FSC, and are transient. The number of different configurations to be
checked increases as $2^{\ell m}$, and this brute-force approach does not
seem to be
very practical for large $\ell, m$. However, experimenting with
rectangular tiles of size $\ell \times m$ leads us to conjecture that
\beq
N_{min}(\ell,m) = f(\ell,m)
\eeq
where
\begin{eqnarray}
f(\ell,m) &=&  \frac{1}{2}(\ell -1)(m-1) , {\rm~~~if}~ \ell {\rm ~or} ~ m
~{\rm odd},
\cr
&=&\frac{1}{2} (\ell m - \ell - m + 2),{\rm ~~~ otherwise}.
\end{eqnarray}

We have checked that the conjecture is true  for $1 \leq \ell, m \leq
4$, and if $\ell \leq 2$ for all $m$. Note that in the limit of large
$\ell, m$, this implies that 
\beq
\rho_{min} = \frac{1}{2}.
\eeq

As additional support for this conjecture, we show that 
$N_{min}(\ell,m)$ cannot exceed $f(\ell,m)$. This is done by
explicitly constructing a recurrent configuration having $f(\ell,m)$ 
particles. We recall the definition that a configuration is
recurrent if it occurs with non-zero probability in the steady state
of the system. The configuration with all sites occupied can clearly 
be reached from any configuration of the system ( by adding  particles
only at unoccupied sites), and is recurrent. Thus a configuration is
recurrent if and
only if there is a non-zero probability that it is reached in a
finite number of steps from the fully occupied configuration.

We now show that for all values of $\ell$ and $m$, starting from a
fully occupied configuration of size $\ell \times m$, we can construct
a sequence of topplings and particle additions that makes the
sandpile reach a
configuration in which the number of
particles
is $f(\ell,m)$. The claim is easily checked for $\ell, m \leq 2$.
For larger values of $\ell,m$, we prove this result by 
induction. Assume that The claim is true for some value $(\ell_0,
m_0)$. We now show that it must then also be true for $(\ell=\ell_0+2,
m_0)$. 

We construct the minimal configuration from the fully
occupied configuration of the $(\ell_0+2,m_0)$ lattice as follows:
 In the initial configuration, $\{z_{x,y}\} =1$, for all $x,y$.
We first produce a configuration on the rectangle $0 \leq x \leq
\ell_0, 0 \leq y \leq m_0$ having only $f(\ell_0,m_0)$ sandgrains. This
can be done, by assumption. In this
process, topplings at the column $x=\ell_0$ would add particles to the
column $x=\ell_0 +1$. We do not relax these yet, and add a particle
to each site of the column $x=\ell_0+2$, and topple it horizontally.
Finally, we relax the sites in the column $x=\ell_0 +1$ by toppling
them in the vertical direction only.
In the  $x= \ell_0+1$ column, we
essentially have a one dimensional ASM with deterministic
topplings. Thus, there can be at most one unoccupied site in the
column after relaxation.
[If all sites in the column are occupied, one vacancy is
created on adding one more particle at any site in the column, and relaxing,
again
transferring particles only in the vertical direction]. The final
number of particles in the configuration is clearly
$f(\ell_0,m_0)+(m-1) = f(\ell_0+2,m_0)$.

By symmetry between the horizontal and vertical directions, we have
$N_{min}(\ell,m) = N_{min}(m,\ell)$. 
Hence, by induction on $\ell$ and $m$, we conclude that
\beq
N_{min}(\ell, m) \leq f(\ell, m), {\rm ~~~for~ ~all} ~\ell,~m.
\eeq
where $f(\ell,m)$ is the integer function defined by eq(20). A proof of
the conjectured equality eq.(18) remains an open problem.

I thank Satya N. Majumdar for a critical reading of the manuscript.

\end{document}